\documentclass[%
 aip,
 amsmath,amssymb,
 reprint,%
]{revtex4-1}

\usepackage{graphicx}
\usepackage{dcolumn}
\usepackage{bm}
\usepackage{placeins}

\usepackage[utf8]{inputenc}
\usepackage[T1]{fontenc}
\usepackage{mathptmx}

\begin{document}

\title{A theory of turbulence mechanics based on material failure}

\author{Samuel J. Raymond}
%

\begin{abstract}
Considerable effort has been expended over the last 2 centuries into explaining the behavior of fluid flow after the onset of turbulence. While perturbations in the velocity field have been shown to explain turbulent transitions, a physical explanation of why flows become turbulent, based on the forces felt by the fluid particles, has remained elusive. In this work a new theory is proposed that attempts to explain the transition of fluid flow from laminar to turbulent as explained by the fluid material undergoing failure. In a vaguely similar sense to how fractures can occur in solids once the balance of momentum exceeds the capacity of the material, so too in a fluid, after sufficient kinetic energy has been achieved by a fluid packet, the viscous forces are unable to maintain the laminar behavior and the fluid packets receive a boost as the stored energy in the viscous bonds are transferred to the kinetic energy of the fluid. This new model is described in terms of fluid packets and the forces on a mass element and commonly-known turbulent flows are used as examples to test the theory. Predicted flow profiles from the theory match the experimental observations of averaged flow profiles and a new equation to predict the onset of turbulence for any flow is presented. This process of the fluid undergoing failure can be seen as a natural continuation of the prevailing wisdom of turbulence when viewed from a different frame of reference.
\end{abstract}

\maketitle

	\section{Introduction}
	\label{sec:intro}
	Observed fluid flows can be categorized into two regimes: laminar and turbulent. Laminar flow is the smooth, linear, viscous dominant flow behavior whereas turbulent flow  is the observed nature of fluid flows that exhibit chaotic, unsteady, motions (\cite{Davidson2015, moffatt_1981, Hussain1975}). The study of turbulent flows by traditional fluid mechanics means has yielded hundreds of subtopics dealing with factors such as boundary layer phenomenon (\cite{LiEwer1985543}), numerical computational efforts (\cite{Launder1974269, Chen1998329, Moin1998539} ), open channel flows (\cite{Kim1987133, Nezu1993}), energy dissipation (\cite{Menter19941598, kolmogorov_1962}),  and different closure models (\cite{Mellor1982851, Reynolds1895}) for statistical approaches. Turbulence has become surprisingly difficult to model with the chaotic, seemingly random motion resisting a clear theory to describe and predict its motion. A theory of turbulent flow has eluded scientists for hundreds of years and remains one of the last true problems in classical mechanics. The vast majority of fluids have been studied, and continue to be studied, in the Eulerian frame of reference, that is, if one were to analyze the flow of a river, a control volume is established in space and the description of the fluid is based in terms of the fields of pressure and velocity as they change within that volume. The particles that make up the fluid flow in an out and are not, explicitly, tracked or used to calculate these variables. Compare this to solid mechanics, where the Eulerian viewpoint is replaced with the Lagrangian, or material view. Here a material volume is used as the building block of the view of the material. Stresses and strains that are imposed upon the material volume are summed over the entire body in question. These two views arise naturally when dealing with solids or fluids as they provide computational efficiency. And while both fluids and solids are governed by the same Newtonian physics underneath, they are represented by two very different forms of governing equations. 
	For fluids, the Navier-Stokes equations arise from performing a balance of Newton’s second Law on the infinitesimal volume of fluid, with the fluxes of quantities like mass and momentum resulting in the final form that includes a notorious, nonlinear term that has been the bane of most fluid mechanists since the equations were first discovered. On the other hand, for solids, a different equation is used. Applying Newton’s second law to an infinitesimal mass volume, and remaining in the Lagrangian, or material, viewpoint the Cauchy equations of motion are derived, with an absence of nonlinear velocity terms. The trade-off here is that while in the Eulerian view the computation of a large mass of fluid is relatively simple, for a solid these mass elements must all be included in the solution, each point must be tracked in time and its evolution calculated.
	The aim of the work presented herein is to formulate a new approach to understand the mechanics of turbulence. This is done by first assuming that a control volume can be approximated by a collection of mass elements and that the stresses that are used to describe fluid flows in conventional fluid mechanics can be  instead framed as forces acting on a collection of fluid masses. Turbulence then becomes a change in the balance of these forces. More specifically, turbulence is introduced here as a failure mechanism whereby the cohesive, viscous, bonds that act between neighboring points in a fluid is broken, and the stored energy of this bond is transferred to the kinetic energy of the mass element, resulting in the momentum-dominant behavior that turbulent flows are characterized by. To introduce this new theory, the paper is structured as follows: firstly the concept of the control mass of a fluid element, and how that relates to the control volume of a fluid is discussed, with the consequences of doing so addressed. Then the forces that are present on a fluid mass element are identified and an attempt to codify them based on notions of solid mechanics and conventional fluid mechanics is performed. Next a form of the mechanical energy of these fluid particles/packets is quantified. Then the notion of the mechanics of turbulence as failure is invoked whereby the kinetic energy receives a boost and the stored viscous energy is released. Finally some examples of common fluid flows, such as pipe flow and plate driven flows are used to calculate the predicted turbulent velocity profiles. A discussion of the results of this approach concludes the paper addressing the new concepts and new questions that this theory poses.

	\section{Building a theory of turbulence mechanics}
	\label{sec:meth}
	In order to show the energy-based model of turbulence as a failure mechanism, first a few concepts need to be introduced and structured properly to illustrate the physics at play. This begins with the idea of moving from a control volume of fluid to a control mass of a fluid packet, then to the identification of the forces on those fluid packets, and finally an expression of the energy of a fluid packet and the consequences of changes to the partitioning of energy during the onset of turbulent flow.
	\subsection{Eulerian Control Volume and the Notion of Fluid Control Mass}
\begin{figure}
		\includegraphics[width=\linewidth]{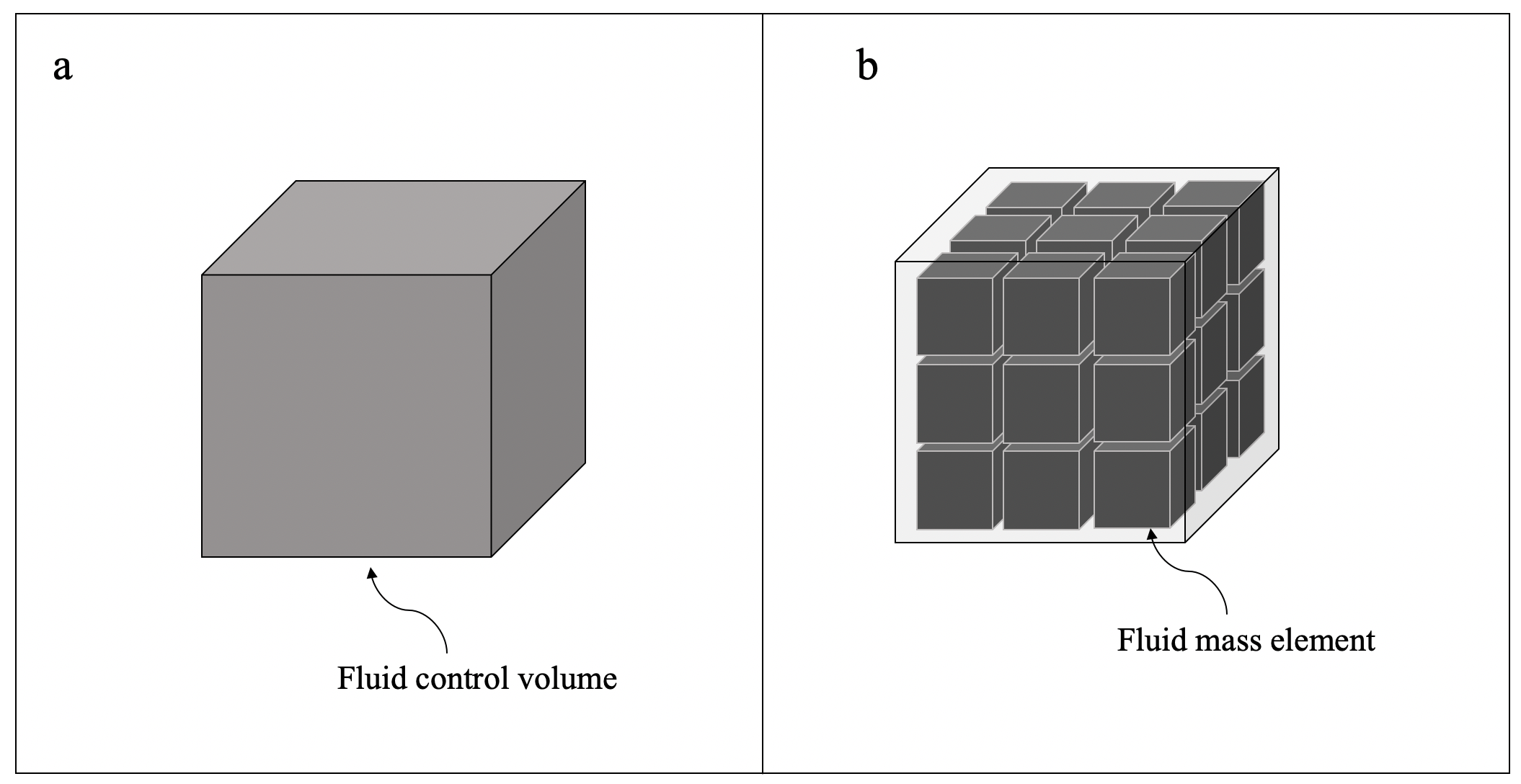}
		\caption{A control volume (a) of fluid is the basis for the derivation of the Navier-Stokes equations. This can also be thought of as a collection of fluid mass elements (b) }
		\label{fig:cvtocm}
	\end{figure}

	A key building block in the equations that are used to describe fluid flow, the Navier-Stokes equations, is the notion of a control volume (\figurename~\ref{fig:cvtocm}a). A control volume consists of a number of variables that are used to build these equations. These variable include, primarily, a density and fluxes of momenta, or stresses that act on the surfaces of this volume. The idea of a control mass, on the other hand, can be thought of as the building block of these control volumes( \figurename  ~\ref{fig:cvtocm}b). The density of the control volume can then be thought of as simply the sum of the control masses divided by the number of masses in that volume. The stresses at the surfaces of the control volume, likewise, could be thought of as the averaged forces projected over an area defined by the control volume's surfaces areas. If the frame of reference can be shifted from an Eulerian/spatial view, where the fluid flows through some space, to a Lagrangian/material view where we can imagine sitting on the fluid mass and tracking its motion then the concept of turbulence becomes something that the fluid mass experiences as a sudden change in its behavior based on the changes of the forces acting on that mass.
	\subsection{Forces on a Fluid Control Mass}
	If we consider a particle/packet of fluid as a mass point subjected to forces consistent with the conventional forces in fluid mechanics, there are two obvious notions that need to be accounted for: a viscous force and a pressure force.
	\begin{figure}
		\includegraphics[width=\linewidth]{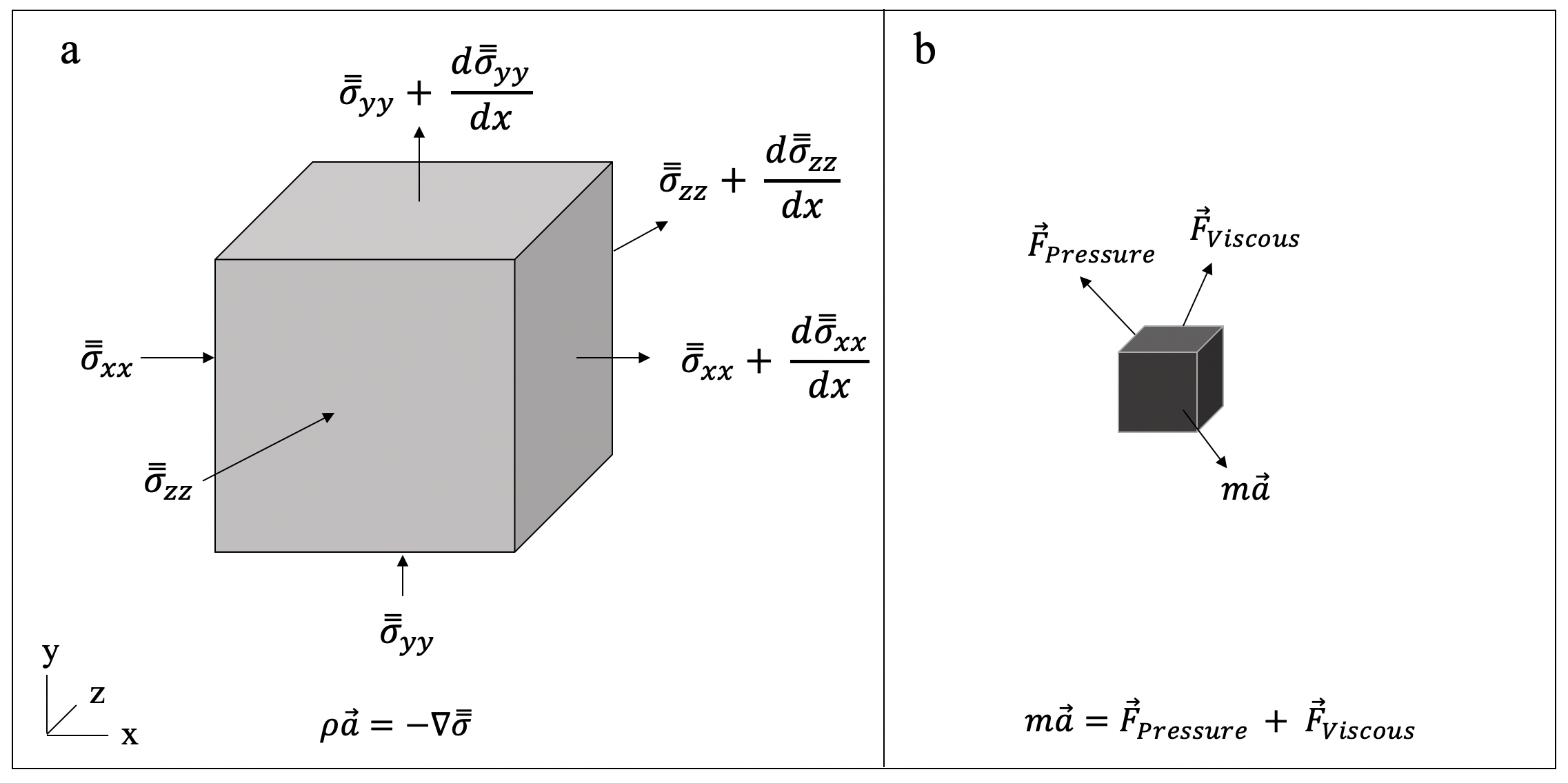}
		\caption{Invoking Newton's 2nd Law on the a) control volume of fluid leads to the derivation of the Navier-Stokes equations and on the b) mass element provides a ways of describing the forces on a fluid mass element.}
		\label{fig:nstosumf}
	\end{figure}
	These two forces can be interpreted from their Eulerian counterparts in the following ways:
	\subsubsection{The Viscous Force}
	Suppose that the fluid is comprised of small fluid mass packets, with interacting forces of attraction that vanish when relative velocities are zero, and that should diminish with inter-packet distance, the following can be proposed:
	\begin{equation}
		F_{viscous}~~\propto~~\frac{m_av_a - m_bv_b}{r^n_{ab}}~~\propto~~\frac{p_{ab}}{r^n_{ab}}
	\end{equation}
	As we are dealing with mass elements, this relative velocity is instead a relative momentum of the neighboring elements. The inverse relationship requires a choice of “n”. Soon we will see that an inverse cube law is a reasonable starting point and will suit our model best. Finally, to ensure dimensional consistency, a coefficient, similar to the viscosity coefficient is required. This coefficient would need to have the units of a cubed length and an inverse time. If we then sum the contributions of these mass elements attracted to their neighbors we arrive at a new expression of the viscous force felt by a fluid packet. 
	\begin{equation}
		F_{viscous} = \sum_{a \ne b}^{N_{ab}} \alpha_{ab}  \frac{p_{ab}}{r^3_{ab}}
	\end{equation}
	The factor $\alpha_{ab}$ has units of a length cubed, $L_{ab}^3$ divided by a time parameter, $T_{ab}$. The material length parameters, $L_{ab}$, can be thought of as the radius that the attractive forces need to be counted for that material, which could be different for fluid packets of different materials. The material time parameter, $T_{ab}$, is less clear as to its obvious connection. But as this parameter is related to the emergent viscosity of the fluid, a shorter time corresponds to a more viscous material, conversely, a shorter length parameter, $L_{ab}$, corresponds to a less viscous material. This form of the viscous force can also be derived by looking at the more conventional shear stress equation for a Newtonian Fluid:
	\begin{equation}
		\tau = \mu \frac{du}{dy}
	\end{equation}
	Where $\tau$ is the shear stress, $\mu$ is the viscosity, and $\frac{du}{dy}$ is the derivative of the x-velocity with respect to the y-direction. By taking the notion of moving from a control volume to a control mass, we can take the following steps. Firstly, let us assume that the shear stress, $\tau$, on the top surface, $A_{\partial \Omega}$, of the control volume (\figurename~\ref{fig:nstosumf}) is the sum of the contributions of the viscous forces, $F_{viscous}$, at the surface:
	\begin{equation}
		\tau = \frac{1}{A_{\partial \Omega}}\int \int F_{viscous} dx dy
	\end{equation}
	Then we can differentiate and rearrange the above to find an expression for the viscous force:
	\begin{equation}
		F_{viscous} = A_{\partial \Omega} \frac{d^2}{dxdy}(\tau)
	\end{equation}
	and if we invoke the expression for the Newtonian shear stress:
	\begin{equation}
		F_{viscous} = A_{\partial \Omega} \frac{d^2}{dxdy}(\frac{du}{dy}) = \frac{d^3}{dxdydy}(u)
	\end{equation}
	The final step is to move from the notion that the velocity is a field, to one where the velocity is actually discrete, so that a derivative of velocity is actually expressed as follows:
	\begin{equation}
		\frac{du}{dx} \approx \frac{u_a - u_b}{r_{ab}}
	\end{equation}
	then we can rewrite the viscous force in terms of discrete velocities and relative positions:
	\begin{equation}
		F_{viscous} \approx  \frac{u_a - u_b}{r_{ab}^3}
	\end{equation}
	which is essentially the form of the equation for viscous forces that was proposed earlier. This conclusion is based on two key propositions: 1) that the fluid can be modeled as a collection of discrete, finite, mass elements, and 2) that the viscous force is proportionate to the relative velocities of neighboring mass elements.\\
	That this force is in the form of an inverse-cube law is potentially quite revealing in the sense of fluid behavior. Inverse cube laws have solutions that correspond to spiraling trajectories (known as Cotes' spirals  - \cite{cotes1722harmonia}) and could therefore be the underlying origins of why eddies and other vortex-like structures are so prevalent in fluids and, in particular, during viscous-relevant processes like energy dissipation. 
	
	\subsubsection{The Fluid Pressure Force}
	From the perspective of the fluid packet/particle, in addition to the viscous attraction force described above, an equilibrium force persists that maintains a stable density of particles in a given volume of fluid. This force can be seen as the fluid-state expression of the elastic equilibrium forces that maintain the isotropic forces in a solid. This force would have an equilibrium distance for two neighboring particles, $d_{ab}$, for some material and when fluids packets concentrate and the relative distances reduce, positive, repulsion forces act upon the fluid packets, likewise when fluid packets are further apart, the attractive forces would result. This is the Lagrangian corollary of the traditional pressure field that are fundamental for fluid mechanics. Pressure differentials are a key component in fluid flows and as many pressure equations are based on a variation in the density field, this can be seen as few fluid packets in a given area, with larger distances between them. This force is then also a force built up by neighbor contributions.
	\begin{equation}
		F_{pressure} = \sum_{b\ne a}^{N_{ab}} K_{ab}(d_{ab} - r_{ab})
	\end{equation}
	Here $d_{ab}$ is an equilibrium distance between the fluid packets ‘a’ and ‘b’ and $r_{ab}$ is the distance between the two packets. For dimensional consistency, a material constant, $K_{ab}$, would be required which would serve a similar role to the stiffness coefficient in solids.
	\subsection{Energy of a Fluid Packet}
	With new expressions for the forces on a fluid packet, we can construct the mechanical energy of a fluid packet, assuming that the forces described above come from a potential energy field. For a typical particle in mechanics, the energy of that particle is the sum of its kinetic and potential energies:
	\begin{equation}
		E_a = \frac{1}{2}m_a v_a^2 + V_a
	\end{equation}
	where $E_a$ is the total mechanical energy of particle 'a', $m_a$ is the mass of that particle, $v_a$ is the velocity of that particle, and $V_a$ is the potential energy of the particle. With the expressions for the viscous and pressure forces, the potential energy would then be:
	\begin{equation}
		V_a = \sum_{a \ne b}^{N_{ab}} \alpha_{ab}  \frac{p_{ab}}{r^2_{ab}} + K_{ab}r_{ab}(d_{ab} - r_{ab})
	\end{equation}
	where the force is derived from this potential energy as: $F = -\frac{dV}{dr}$. We can then express the mechanical energy combining the kinetic and viscous energies.
	\begin{equation}
		E_a = \frac{1}{2}m_a v_a^2 + \sum_{a \ne b}^{N_{ab}}( \alpha_{ab}  \frac{p_{ab}}{r^2_{ab}} + K_{ab}r_{ab}(d_{ab} - r_{ab}) )
	\end{equation}
	\subsection{Turbulence as a Failure Mode}
	Now that we have an expression for the energy of a fluid packet, and a notion that the fluid material can be comprised of interacting fluid masses, we can introduce the idea of material failure. If we return to the idea of traveling with a fluid packet, we can ask the question of what we would see when laminar flow becomes turbulent, in this new frame. For a solid, failure occurs when the elastic attraction forces between neighboring masses is overcome by the momentum of the masses and the stored energy is released as plastic flow, fracture, heat, sound etc. 
	\begin{figure}
		\centering
		\includegraphics[width=\linewidth]{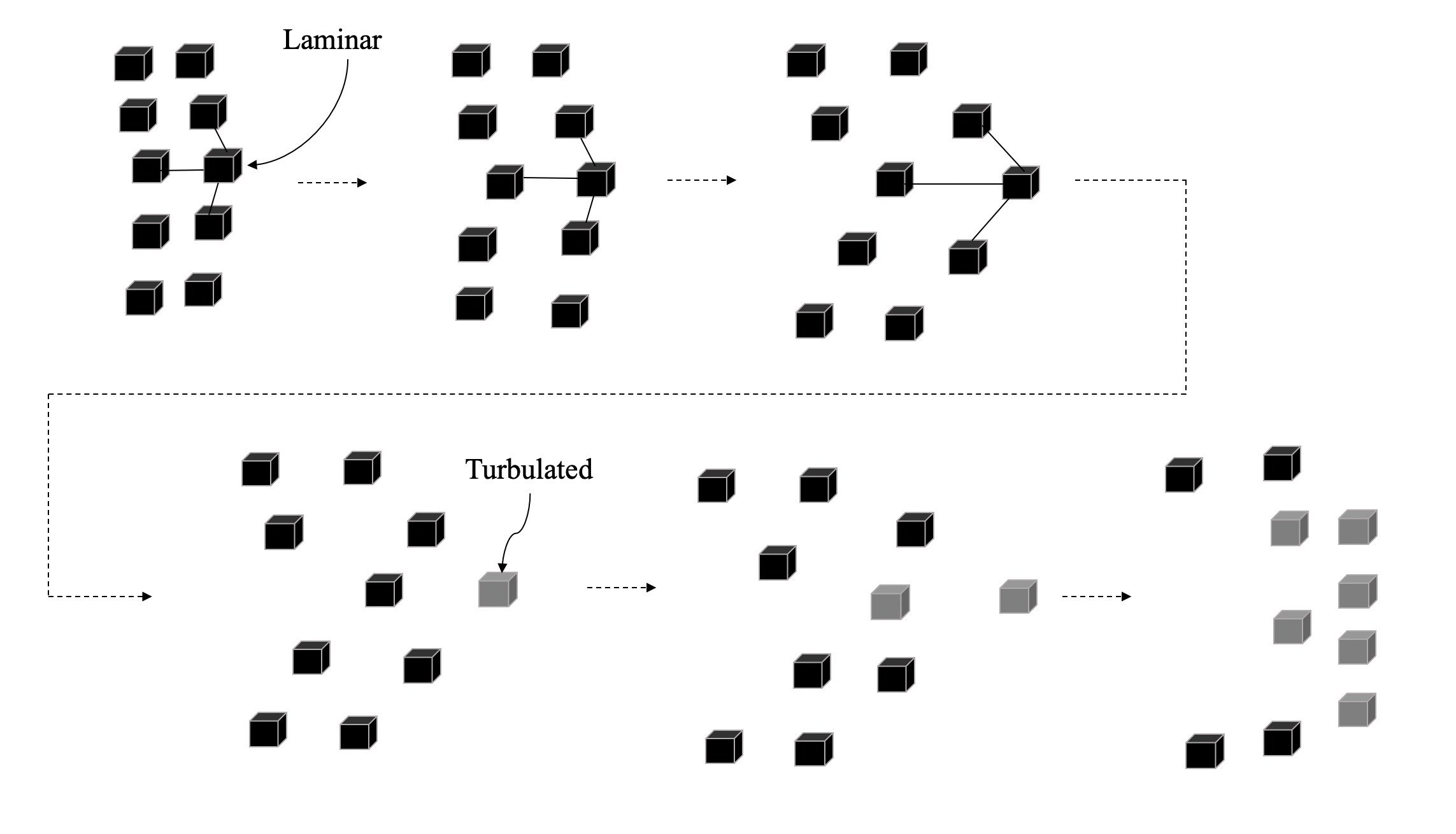}
		\caption{Turbulation occurs when the viscous force bonds are broken from the surrounding neighbor fluid packets and the turbulated packet is given a boost in velocity, this process cascades are the flow gets faster and turbulent flow is achieved.}
		\label{fig:turbulation}
	\end{figure}
	For a fluid, this attractive force is the viscous force that prevents two masses from increasing their relative velocities. If the relative velocity between two neighbors is too large, the stored energy in the viscous attraction is released and the fluid packet receives a boost in its kinetic energy as it is no longer being held back by its slower neighbor, a schematic of this is shown in \figurename ~\ref{fig:turbulation} . If we look at the energy equation from the previous section, and for simplicity ignore the energy stored in the pressure force, we can see the result of this failure process. If the energy stored in the viscous force is transferred to the kinetic energy of the fluid packet we would find the following. The energy for the laminar flow, $E_a^{Laminar}$, 
	\begin{equation}
		E_a^{Laminar} = \frac{1}{2}m_a v_a^2 + \sum_{a \ne b}^{N_{ab}} \alpha_{ab}  \frac{p_{ab}}{r^2_{ab}}
	\end{equation}
	then if we assume that energy is conserved right at the point of \textit{turbulation} (when the fluid packet fails),
	\begin{equation}
		E_a^{Turbulent} = E_a^{Laminar}
	\end{equation}
	and that now all of the energy is transferred to the kinetic energy of the fluid,
	\begin{equation}
		E_a^{Turbulent} = \frac{1}{2}m_a (v_a^{Turbulent})^2 
	\end{equation}
	we can calculate the velocity of the turbulated fluid packet,
	\begin{equation}
		v_a^{Turbulent} = \sqrt{\frac{2 E_a^{Turbulent}}{m_a} } 
	\end{equation}
	this result indicates that we can, in theory, smoothly track the velocity, and hence the motion of a fluid packet from laminar to turbulent flow. The cost here is in the computational effort of keeping track of all of the points and understanding the interaction between collisions of laminar and turbulated fluid packets. In the next section we will see that this theory predicts the flat velocity profiles that are found in studying fluid flows and should allow for the detailed description and calculation of any turbulent flow. 
	
	\subsection{Equation of Turbulent Failure - Moving Beyond the Reynolds Number}
	One missing piece is determining when the relative velocity will become too large for neighboring fluid packets. Here we would need to find a material-specific equation that predicts the relative velocity after which a fluid packet would turbulate. The Reynolds number gives a good place to start. During the experiments he conducted (\cite{Reynolds1883}),  Reynolds found a dimensionless number that correlated well with low values associated with laminar flow and high values associated with turbulent flow. This number is a ratio of the momentum of the fluid to the viscous forces of the fluid. A larger value indicating that the momentum has exceeded some balanced point and therefore the flow was turbulent. Using these experimental results and the new energy equation for the fluid packet, we can derive a new equation to predict the onset of turbulence. Let us assume there is a turbulence equation $f_a^T$ that is a function of relative positions and velocities and that when this equation equals zero, the fluid packet has failed.
	\begin{equation}
		f_a^T(v_{ab},r_{ab}) = 0
	\end{equation} 
	If we assume that the function is based on energy, and that the viscous energy is the dominant energy relevant to turbulence, then we can propose:
	\begin{equation}
		f_a^T(v_{ab},r_{ab}) = \sum_{a \ne b}^{N_{ab}} \alpha_{ab}  \frac{p_{ab}}{r^2_{ab}}  - E_{crit}
	\end{equation} 
	Where the critical energy, $E_{crit}$, is a material parameter that represents the capacity of the fluid material to maintain viscous forces between fluid packets. \\One criticism of the Reynolds number is that it cannot be universally applied to all fluid flows, nor is it very useful in predicting when a given flow will become turbulent. The characteristic length that is present in the original form causes issues when discussing open channel flow or when the characteristic length choice is unclear. This approach, however, does not show these limitations, it should be applicable to any fluid situation. The downside being having to track the motion of fluid packets and the neighbors.
	
	The similarities here between solids and fluids provide an interesting notion of symmetries across states of matter and their behavior. If one were to study the motion and shape of fractures in solids, they would find themselves in complex mathematical topics like bifurcation, fractal geometry and chaos. These kinds of fields are heavily present in modern turbulence flow studies too, and have shown to be extremely valuable. In both the solid and the fluid case, when trying to understand how these mathematical structures emerge, looking from the point of view of the fluid packet can offer predictive insights, such as when and where failure/turbulence will occur. For solids, fracturing occurs when the internal stresses within a material exceed the elastic limit and the energy of these elastic bonds is transferred either into plastic flow (permanent deformation from the loading) or surface creation (i.e. fractures). A key difference between solids and fluids is the phase transition that has occurred from a solid to a fluid. The elastic forces that solids are bound by are not as restrictive to motion in fluids and allow the flow of material. However, the fluid is still attracted to itself internally, producing the friction between the fluid, commonly referred to as viscosity. Laminar flow often exhibits many similar qualities to elasticity (reversibility, linearity, etc.), but as the fluid packet is freely moving from one set of neighbors to another, the picture of a rigid elastic bond such as those present in solids does not work. Instead a new picture of the bonds between packets of material is needed for this new view of fluids that carries some of the features of elastic bonds in solids but is less restrictive. It is also important to note here that turbulent flows exhibit chaos and unpredictability. While this proposed model allows the prediction of the onset and development of fluid flow in a turbulent regime, it does not prevent the dynamics of the flow from exhibiting a nonlinear, chaotic nature and the other aspects that are well understood of turbulent flows. 
	
	In the next section this new theory will be applied to some well known cases of conventional turbulence studies to see how well these new models are compatible with existing knowledge.
	\section{Applying this new turbulence theory to common fluid flow example}
	\label{sec:results}
	To test the newly proposed theory, we can apply this model to some well known, simple, fluid flow examples. Taking the starting boundary conditions of a well studied problem and then applying the fluid packet approach, we can see that the same results can be derived.  
	\subsection{Pipe Flow}
	The quintessential flow example used to investigate turbulence is the pipe flow example that Reynolds studied in his seminal work. According to the new theory presented here, at the point where a fluid packet transitions from a laminar state to a turbulated state, the viscous energy is transferred to the kinetic energy. For the flow in a pipe, the fully developed flow profile follows a quadratic in the form:
	\begin{equation}
		U(y) = U_0 (1-(\frac{y}{R_0})^2)
	\end{equation}
	\begin{figure}
		\centering
		\includegraphics[width=\linewidth]{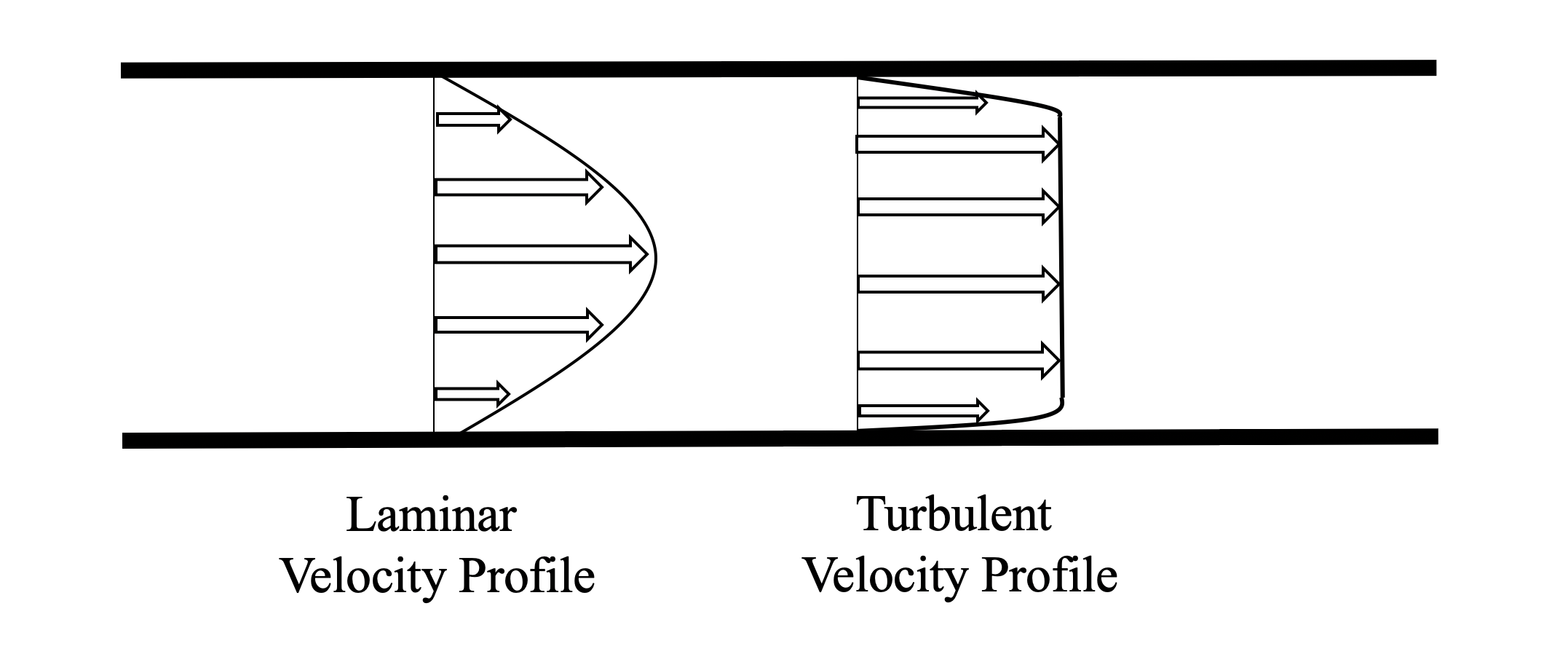}
		\caption{Conventional studies of pipe flow show the average velocity profile transitions from a parabolic form in laminar flow to a flattened form in turbulence where the average velocity is relatively constant across the diameter of the pipe. }
		\label{fig:lamtoturpipe}
	\end{figure}
	
	Where $U(y)$ is the x-velocity as a function of the y-coordinate, $U_0$ is the maximum speed of the flow, and $R_0$ is the radius of the pipe. Once the flow has become turbulent, the averaged velocity field shows a flat profile across the channel diameter. (\figurename ~\ref{fig:lamtoturpipe}).
	We can apply the velocity distribution of the pipe flow example and create a model of fluid packets with velocities according to their y-coordinate. If we then calculate the viscous forces and build the energy of the fluid packets we can then invoke turbulation and look at the velocity profile that is predicted to compare with the laminar case.
	\begin{figure}
		\centering
		\includegraphics[width=\linewidth]{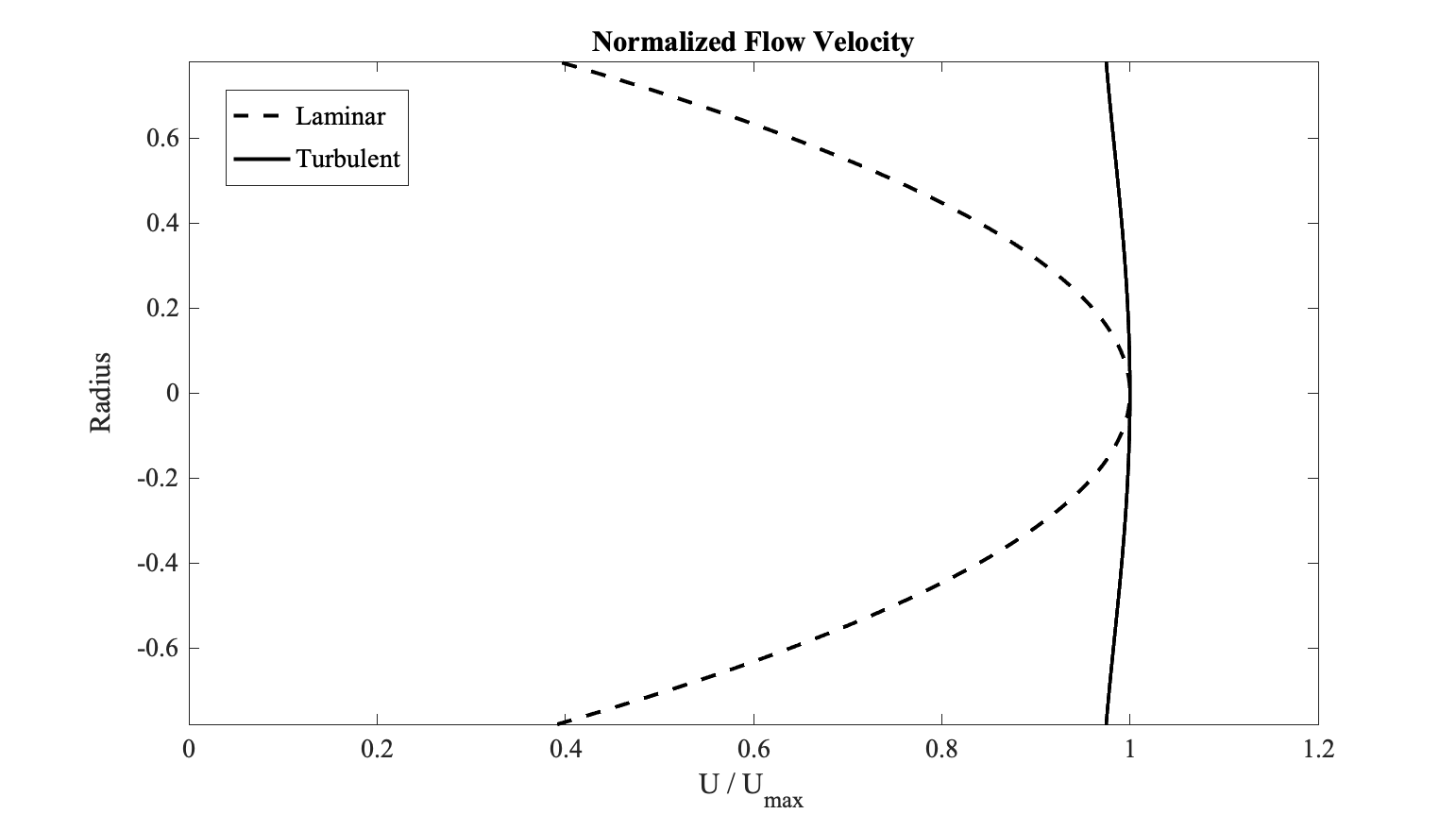}
		\caption{Calculations of the fluid packets using the viscous force model and the notion that the energy of the fluid transitions to be entirely kinetic at turbulation predicts the flat velocity profile seen in experimental data.}
		\label{fig:pipe_calcs}
	\end{figure}
	\figurename~\ref{fig:pipe_calcs} shows the different velocity profiles (normalized to the maximum velocity) for the laminar and turbulent case for pipe flow using the fluid packet description. It is important to note here that this state of a uniform flow field is not expected to arise in flows as the fluid constantly interacts with itself, this result would be the average result of the fluctuations that occur as fluid packets turbulate.
	\subsection{Couette Flow}
\begin{figure}
		\centering
		\includegraphics[width=\linewidth]{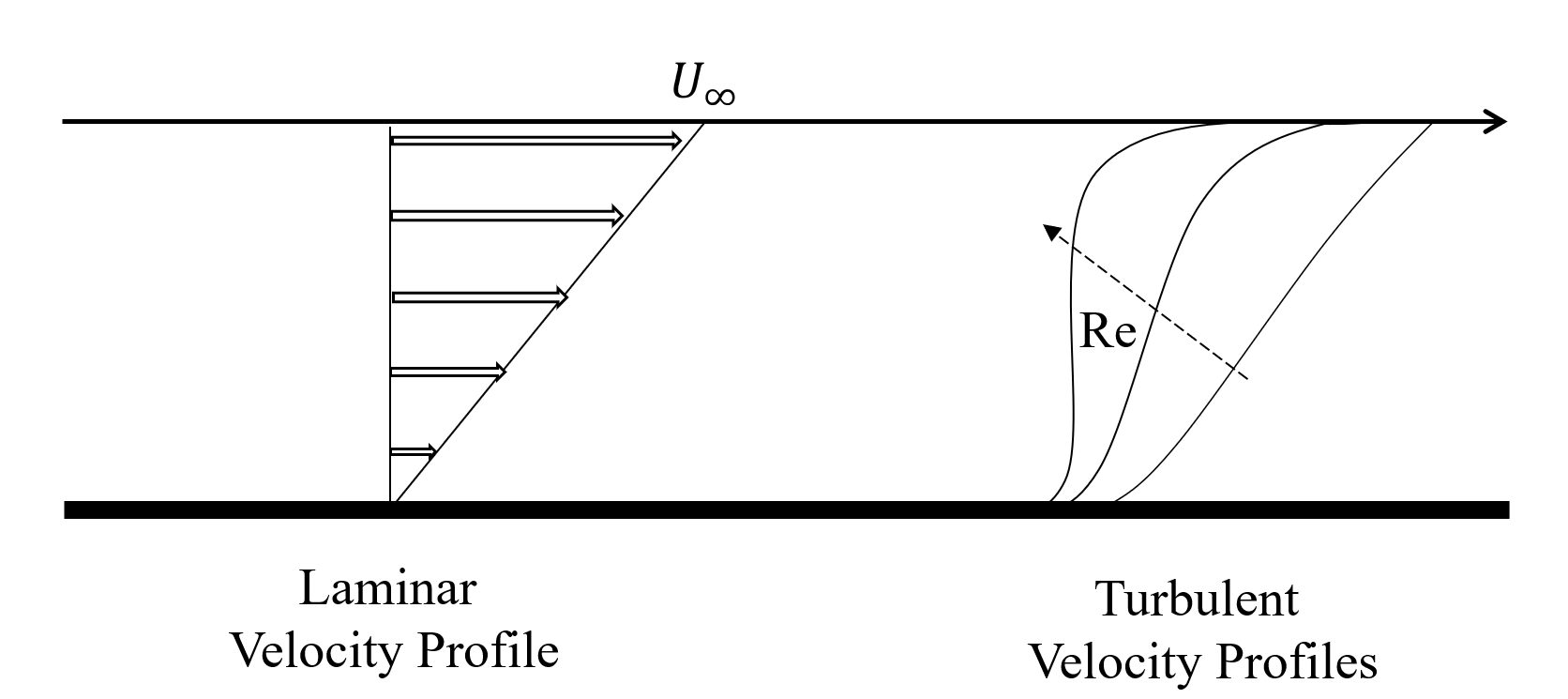}
		\caption{Conventional studies of pipe flow show the average velocity profile transitions from a parabolic form in laminar flow to a flattened form in turbulence where the average velocity is relatively constant across the diameter of the pipe. }
		\label{fig:lamtoturcouette}
	\end{figure}

	For another example, consider the Couette flow that results from a plate driven flow. The velocity profile for fully developed, laminar flow, is:
	\begin{equation}
		U(y) = U_{\infty}\frac{y}{h}
	\end{equation}
	where $U(y)$ is the x-velocity as a function of y-coordinate, $U_{\infty}$ is the plate velocity, and $h$ is the height of the plate above the stationary surface. The velocity of the turbulent flow has been measured at different Reynolds numbers to behave as shown in \figurename~\ref{fig:lamtoturcouette} with the sloped velocity profile gradually becoming steeper with increased Reynolds number.
	\begin{figure}
		\centering
		\includegraphics[width=\linewidth]{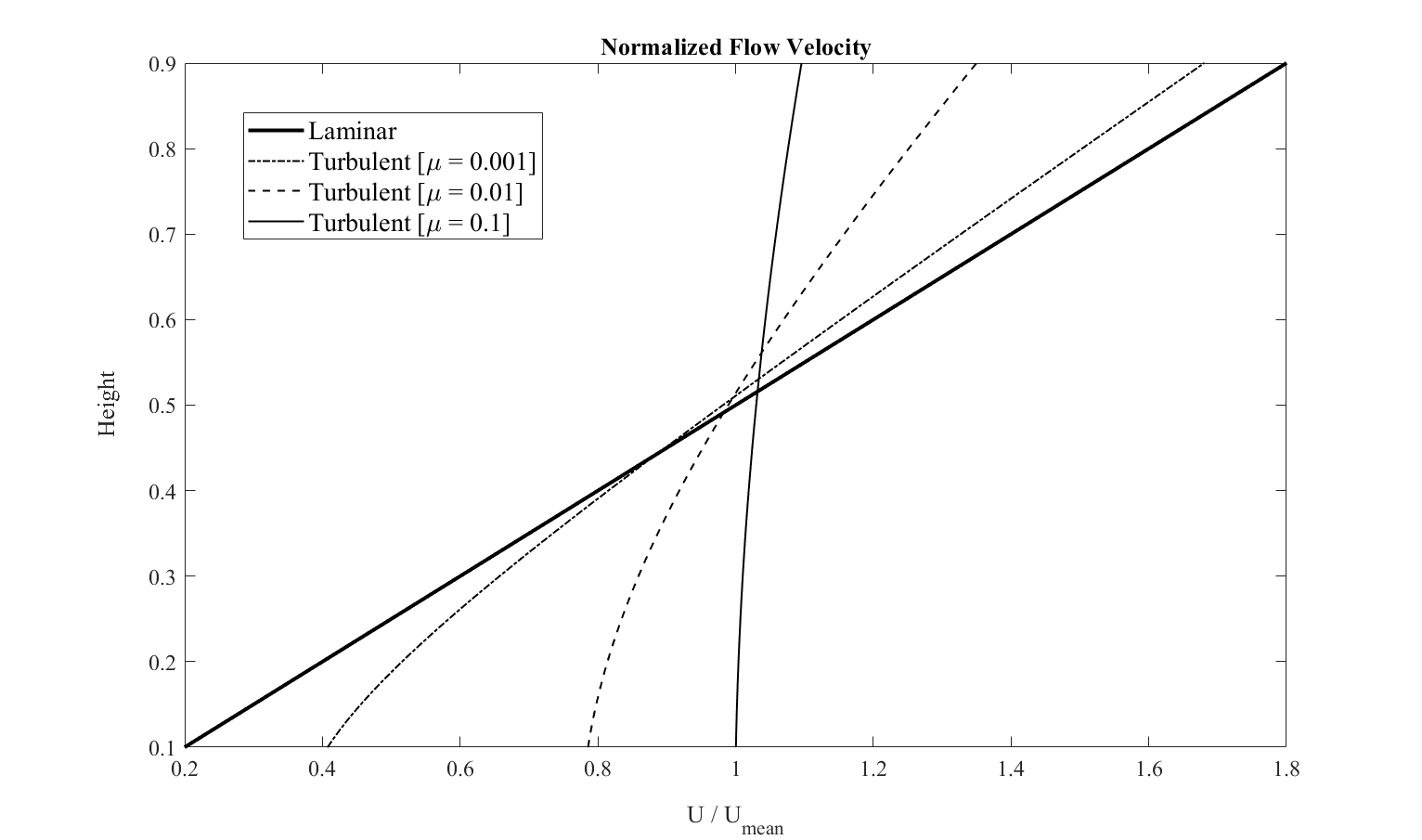}
		\caption{Velocity profiles in the plate driven Couette flow of the laminar and turbulent cases for larger Reynolds numbers, indicated by larger viscosity coefficients $\mu$. As the Reynolds number increases for a given turbulent flow, the steepness of the profile is predicted to increase, which is in line with experimental observations.}
		\label{fig:coucalcs}
	\end{figure}
	By seeding the velocities of a series of fluid packets with the velocity distribution for a fully developed flow profile, we can apply the same process as for the pipe flow example. However in this case, to observe the effect of higher Reynolds number flow we can use different viscosity coefficients. Three values, 0.1, 0.01, and 0.001 for the sake of illustration are used and the resulting turbulent velocity profiles are compared in \figurename~\ref{fig:coucalcs}. Here we can see the steepening turbulent velocity profiles resulting in more stored viscous energy being transformed to kinetic energy during turbulation relative to the laminar flow regime. 
	These results give a strong indication that the viscous force is a key to turbulence mechanics and suggesting turbulence as a failure mode of fluids gives a direction to further investigation of flows.
	\FloatBarrier
	\section{Discussion}
	\label{sec:disc}
	A new theory has been proposed that seeks to explain the physical mechanism that gives rise to turbulent flow in fluids. This theory states that a fluid can be thought of as a collection of fluid packets, and that these packets interact with their neighbors by two forces, a viscous force and a pressure force. The pressure force is reminiscent of the elastic force in solids and represents an equilibrium spacing of fluid packets with respect to their neighbor such that a concentration of neighbors results in a large, positive pressure and a dearth of neighbors resulting in a negative pressure force. The viscous force is theorized to be in inverse cube law force proportional to the relative velocities between neighboring fluid packets. The theory states that at the point where the relative velocities of two neighboring fluid packets reaches a critical point, turbulation will occur, the viscous bond will be broken and the stored energy will be transferred to the fluid packets' kinetic energy, resulting in the turbulent behavior seen in fluid flows. This theory  necessitates that the fluid be modeled not in terms of the traditional control volume, but instead a collection of control mass elements, or fluid packets. \\
	Whether or not these finite mass elements are a fundamental feature of the fluid, such as a molecule of the material, or more conceptual, is still unknown at present. This does have some interesting consequences for the Kolmogorov scales  (\cite{kolmogorov_1962}) and dissipation of turbulent energies as the smallest conceivable eddy in this model would be two molecules rotating around a central point with the viscous forces causing them to rotate in a spiral. The dissipation of energy (\cite{richardson_lynch_2007}) and the transition of turbulent back to laminar has not been covered in this work and is still an unresolved matter. \\
	Additionally, while turbulence was the focus of this theory, using the fluid packet view of fluids and the mechanical energy of said fluid packet, there are additional failure modes that can be theorized in the same model. The first is one that occurs when dealing with the surface of a fluid. Here we arrive at a situation that more closely mirrors the situation for solids. Surface energy at the boundary of two fluids like air and water becomes as important as the surface energy of a solid as a brittle fracture develops. Consider the example of a leaky faucet, the fluid can be seen as a small volume with a large surface area, under the influence of gravity. As the fluid is accelerated, the material strength of the viscous bonds is overcome by the increased momenta, as there is no more material around the small volume of water, the energy from the viscous bond is transferred to surface energy and the droplet closes off and drips down. This would require additional terms in the energy and failure equation, but conceptually still fits in the proposed model. Finally, the second additional failure mode that can be described with this model occurs in extreme conditions that give rise to cavitation. This is somewhat similar to the droplet formation but occurs within the volume of the fluid. As the isotropic pressure increases, the fluid is forced apart further and further until the energy to create a new surface is lower than the stored viscous energy and a cavitation bubble is created. Both of these different failure modes require more investigation\\
	
	While this newly proposed theory takes an alternate frame of view to conventional fluid mechanics, there are still unresolved matters that this current work has not addressed. A full, formal, 3-dimensional mathematical description of the forces at play, their validation with accepted theoretical and experimental work is still needed. As is a description of the boundary layer and the role that these forces play as the boundary layer changes and laminar fluid packets interact with turbulated ones. The dissipation of turbulent energy and the recovery to laminar flow is also still an open question. The hope is that this new approach and the examples given will provide the critical impetus to exploring further into this alternative view of fluids and turbulence. 
	\section{Conclusion}
	\label{sec:conc}
	A new theory of turbulence is proposed that suggests the fundamental mechanism for turbulence is material failure. This is based on the assumption that a fluid can be modeled as a collection of fluid mass elements, each interacting with neighboring elements via two dominant forces, the pressure force and viscous forces, forms of which are presented. Turbulence is the scenario that occurs when the relative velocity between two fluid packets exceeds the critical material limit and the stored viscous energy is transferred to the kinetic energy, resulting in the jumps in momenta that accompany turbulent flows. Examples are used for traditional fluid flow problems in pipes and plate driven scenarios and the new model is able to predict the behavior of the fluid velocity fields after the onset of turbulence.  This allows the smooth, constant tracking of a fluid velocity, position, and forces both during laminar and turbulent flows as viewed from the perspective of a collection of mass elements. If correct, this new model of fluids could help to gain further insights into the mechanics of turbulence, fluid flows in general, and improve our understanding of matter and energy.
	
	\section{Data Availability}
	Data sharing is not applicable to this article as no new data were created or analyzed in this study.
	\bibliography{turb_bib}

\end{document}